\documentclass[epj]{svjour}
\usepackage{graphics}
\usepackage[T1]{fontenc}
\usepackage[latin1]{inputenc}
\usepackage{amsmath}
\usepackage{graphicx}
\usepackage{amssymb}
\usepackage{color}
\usepackage{graphicx}
\usepackage{epsfig}
\usepackage{multirow}
\usepackage{dcolumn}
\usepackage{bm}
\usepackage{subfigure}

\def\<{\langle}
\def\>{\rangle}

\begin{document}

\title{Vibration induced memory effects and switching in ac-driven molecular nanojunctions}
\author{Andrea Donarini, Abdullah Yar, \and Milena Grifoni
}

\institute{Institute of Theoretical Physics, University of Regensburg, D-93040 Regensburg, Germany}
\date{Received: date / Revised version: date}
%
\abstract{
We investigate bistability and memory effects in a molecular junction weakly coupled to metallic leads with the latter being subject to an adiabatic periodic change of the bias voltage. The system is described by a simple Anderson-Holstein model and its dynamics is calculated via a master equation approach. The controlled electrical switching between the many-body states of the system is achieved due to polaron shift and Franck-Condon blockade in the presence of strong electron-vibron interaction. Particular emphasis is given to the role played by the excited vibronic states in the bistability and hysteretic switching dynamics as a function of the voltage sweeping rates. In general, both the occupation probabilities of the vibronic states and the associated vibron energy show hysteretic behaviour for driving frequencies in a range set by the minimum and maximum lifetimes of the system. The consequences on the transport properties for various driving frequencies and in the limit of DC-bias are also investigated.
} 
\authorrunning{Andrea Donarini, Abdullah Yar, and Milena Grifoni}
\titlerunning{Vibration induced memory effects and switching in driven...}
\maketitle

\section{Introduction}
Quantum switching, bistability and memory effects provide potential applications for molecular electronics~\cite{96APL262107,285Science391,408Nature67,80SA100}.
Recent scanning-tunneling microscopy (STM) experiments~\cite{305Science493,317Science1203,98PRL176803,68PRB045302,97PRL216103}
have shown bistability and multistability of neutral and charged states. Random and controlled switching of single molecules~\cite{98PRL176807,2nanotch176,2small973}, as well as conformational memory
effects~\cite{317Science1203,97PRL216103,96PRL156106,292Science2303} have been recently investigated.
Other groups have observed memory effects in graphene~\cite{4ACSNano7221,5ACSNano7812,99APL113112}
and carbon nanotubes~\cite{325Science1103,5Nature321,3Nanotech26}.
Motivated by the experimental achievements, several groups~\cite{10NJP085002,81PRB115333,5Nanoletter125,67PRB235312,67PRB075301,78PRB085409,73PRB035104}
have attempted to theoretically explain these striking features invoking a strong electron-vibron coupling.
In Ref.~\cite{10NJP085002} charge-memory effects have been investigated in a polaron-modeled system using the equation-of-motion method for the Green's functions in the strong tunnel coupling regime. Similarly, in Ref.~\cite{5Nanoletter125} these effects are associated with a polaron system treated within a simple mean-field approach.
However, the hysteresis effects in Ref.~\cite{5Nanoletter125} may be an artefact of the mean-field approximation as pointed out by Alexandrov and Bratkovsky~\cite{arXiv:0603467v3}.
In  Ref.~\cite{67PRB235312} memory effects have been found in a polaron-modeled system taking the quantum dot as a d-fold-degenerate energy level weakly coupled to the leads and accounting for attractive electron-electron interactions.  However, here a multiple degenerate energy level (d>2) is required.
In contrast, in Ref.~\cite{78PRB085409}, again the situation of weak coupling to the leads but with repulsive electron-electron interaction is considered. In this work, bistability, charge-memory effects and switching between charged and neutral states of a molecular junction have been explained within the framework of a polaron model, where an electronic state is coupled to a single vibronic mode. These features have been associated with the asymmetric voltage drop across the junction and the interplay between time scales of voltage sweeping and quantum switching rates between metastable states in the strong electron-vibron coupling regime.
In the weak tunnel coupling limit, a perturbation theory in the tunneling amplitude between the molecule and leads
is appropriate to describe electronic transport. In particular, such a perturbative treatment is valid if the
tunneling-induced level width $\hbar{\rm{\Gamma}}$ is small enough compared to the thermal energy $k_{\text{B}}T$. The lowest order in this expansion leads to sequential tunneling, which corresponds to the incoherent transfer of a single electron from a lead onto the molecule or vice versa. Moreover, it is known from transport theory that sequential tunneling is dominant as long as the dot electrochemical potential (i.e. the difference $E_{N} - E_{N-1}$ between eigenvalues of the many-body Hamiltonian corresponding to  states with particle number differing by unity) is located between the Fermi energies of the leads.\\\indent
A strong electron-vibron coupling can in turn qualitatively affect the sequential tunneling dynamics~\cite{74PRB205438,94PRL206804,95PRL056801,Nowack0506552,84PRB115432,schultz2010}. For strong coupling, the displacements of the potential surfaces for the molecule in a charged or neutral configuration are large compared to the quantum fluctuations of the nuclear configuration in the vibrational ground state. As a result, the overlap between low-lying vibronic states is exponentially small. This leads to a low-bias suppression of the sequential transport known as Franck-Condon (FC) blockade, which in turn is responsible for bistability effects in~\cite{78PRB085409}.\\\indent
In this paper we extend and improve the ideas of Ref.~\cite{78PRB085409}. Specifically, we include the time dependence of the bias voltage explicitly, and derive a time-dependent master equation for the reduced density matrix of a single level molecule coupled to a vibrational mode and weakly coupled to metallic leads. Moreover, we relax the assumption of fast relaxation of vibrons into their ground states and discuss the role played by the vibronic excited states in the switching dynamics. As in Ref.~\cite{78PRB085409}, we find that controlled electrical switching between metastable states is achieved due to polaron shift and Franck-Condon blockade in the presence of strong electron-vibron interaction. Moreover, we find that the hysteresis effects can be observed in the switching dynamics only if the time scale of variation of the external perturbation, $T_\text{ex}$, is constrained into a specific range set by the minimum, $\tau_\text{min}$, and maximum, $\tau_\text{max}$, charge lifetimes of the system as a function of the applied bias. With $\lambda$ being the dimensionless electron-vibron coupling, it holds $\tau_\text{min}\sim{\rm{\Gamma}}^{-1}$, $\tau_\text{max}\sim{\rm{\Gamma}}^{-1}e^{\lambda^2}$. Hence, a strong electron-vibron coupling ($\lambda\gtrsim 1$) is a necessary condition for the opening of this time scale window and thus of hysteresis. Such a large dimensionless electron-vibron coupling is not rare in conjugated molecules with soft torsional modes (e.g biphenyl with different substituents, azobenzene) which have been experimentally proven to behave as conformational switches ( \cite{2small973}, \cite{96PRL156106}). Very large reorganization energies (of the order of 1 eV) attributed to a polaron effect have also been observed in STM single atom switching devices \cite{305Science493}. Also in this case the electron-phonon coupling should be large ($\lambda \gtrsim 1$) to justify the bistability. Outside this range the averaging over multiple charging events in the slow driving case or multiple driving cycles in the fast case removes the hysteresis.\\
\indent The paper is organized as follows: In Section~\ref{sec:ModelHamiltonian} the model Hamiltonian of a single level molecule coupled to a vibronic mode is introduced. A polaron transformation is employed to decouple the electron-vibron interaction Hamiltonian and obtain the spectrum of the system.\\\indent
In Section~\ref{sec:SequentialTunneling} we derive equations of motion for the reduced density matrix for the case in which the leads are subject to an adiabatic bias sweep. The time-dependent master equation is solved in the limit of weak coupling to the leads and important time scale relations are derived. \\\indent
In Sections~\ref{sec:Lifetimes},~\ref{sec:SwitchingHysteresis} and~\ref{sec:LFC}, our main results of the memory effects are presented and analyzed for a sinusoidal perturbation of period $T_\text{ex}=2\pi/\omega$.\\
In Section~\ref{sec:Lifetimes} the lifetimes of the many-body states of the system are calculated. We show that, for the case of asymmetric voltage drop across the junction, at small bias voltages a bistable configuration is achieved which plays a significant role in the hysteretic dynamics of the system. Bistability can involve also vibronic excited states of the system. \\\indent
In Sections~\ref{sec:SwitchingHysteresis} and~\ref{sec:LFC} we give an explanation of the hysteretic behavior of the system in terms of characteristic time scales, in particular, the interplay between the time scale $T_\text{ex}$ of variation of the external perturbation and of the dynamics of the system set by $\tau_\text{switch}\sim\tau_\text{min}\sim{\rm{\Gamma}}^{-1}$.\\
In Section~\ref{sec:SwitchingHysteresis} focus is on the regime $\omega\sim{\rm{\Gamma}}$ while in Section~\ref{sec:LFC} is $\omega\ll{\rm{\Gamma}}$. In the latter case the features observed in Ref.~\cite{78PRB085409} can be successfully reproduced.\\\indent
In Section~\ref{sec:stationary}, the consequences on the transport properties in the  DC-limit are presented as a special case. Finally, we conclude in Section~\ref{sec:Summary}.

\section{Model Hamiltonian}\label{sec:ModelHamiltonian}
We consider a simple Anderson-Holstein model where the Hamiltonian of the central system is described as
\begin{align}\label{eq:SysHamiltonian}
\hat{H}_\text{sys}=\hat{H}_\text{mol}+\hat{H}_\text{v}+\hat{H}_\text{e-v},
\end{align}
where $\hat{H}_\text{mol}$ represents a spinless single molecular level modeled by the Hamiltonian
\begin{align}\label{SysHamiltonian1}
\hat{H}_\text{mol}=\left(\varepsilon_0+eV_\text{g}\right)  \hat{d}^\dag\hat{d},
\end{align}
where $\hat{d}^\dag(\hat{d})$ is the creation (annihilation) operator of an electron on the molecule and $\varepsilon_0$ is the energy of the molecular level, and $V_\text{g}$ accounts for an externally applied gate voltage. For simplicity we assume a spinless state describing the molecular level with strong Coulomb interaction where only one excess electron is taken into account. The spin degeneracy would not qualitatively change the results of the paper.
The vibron Hamiltonian can be written as
\begin{align}\label{SysHamiltonian2}
\hat{H}_\text{v}=\hbar\omega_0\left(\hat{a}^\dag\hat{a}+\frac{1}{2}\right),
\end{align}
where $\hat{a}^\dag(\hat{a})$ creates (annihilates) a vibron with energy $\hbar\omega_0$.
Finally, the electron-vibron interaction Hamiltonian is expressed as
\begin{align}\label{SysHamiltonian3}
\hat{H}_\text{e-v}=g\hat{d}^\dag\hat{d}\left(\hat{a}^\dag+\hat{a}\right),
\end{align}
where $g$ is a coupling constant.

\subsection{Polaron transformation}\label{sec:PolaronTransformation}
In order to decouple the electron-vibron interaction Hamiltonian, we apply the canonical polaron unitary transformation~\cite{mahan2000}. Explicitly, we set $\tilde{\hat{H}}_\text{sys}=e^{\hat{S}}\hat{H}_\text{sys}e^{-\hat{S}}$, where
\begin{align}\label{polaronoperator}
\hat{S}=\lambda \hat{d}^\dag\hat{d}\left(\hat{a}^\dag-\hat{a}\right),
\end{align}
with $\lambda=\frac{g}{\hbar\omega_0}$ as the dimensionless coupling constant.
The transformed form of the electron operator is
\begin{align}\label{transoperator}
 \tilde{\hat{d}}=\hat{d} \hat{X},
\end{align}
where $\hat{X}=\exp\left[-\lambda\left(\hat{a}^\dag-\hat{a}\right)\right]$. In a similar way, the vibron operator is transformed as
\begin{align}\label{transviboperator}
\tilde{\hat{a}}=\hat{a}-\lambda \hat{d}^\dag\hat{d}.
\end{align}
Now the transformed form of the system Hamiltonian reads
\begin{align}\label{DecSysHamiltonian}
\tilde{\hat{H}}_\text{sys}= \varepsilon\hat{d}^\dag\hat{d}+\hbar\omega_0\left(\hat{a}^\dag\hat{a}+\frac{1}{2}\right),
\end{align}
where $\varepsilon=\varepsilon_0+ eV_\text{g}-\frac{g^2}{\hbar\omega_0}$ is the polaron energy with polaron shift $\varepsilon_\text{p}=\frac{g^2}{\hbar\omega_0}$.
The polaron eigenstates of the system are
\begin{align}\label{eq:eigenstates}
|n,m\rangle_1:=e^{-\hat{S}}|n,m\rangle,
\end{align}
where $n$ denotes the number of electrons on the molecular quantum dot, while the quantum number $m$ characterizes a vibrational excitation induced by the electron transfer to or from the dot.

\section{Sequential tunneling}\label{sec:SequentialTunneling}
We analyze the transport properties of the system in the limit of weak coupling to the leads. The Hamiltonian of the full system is expressed as
\begin{align}\label{eq:FullSysHamiltonian}
\hat{H}(t)= \hat{H}_\text{sys}+\hat{H}_\text{T}+\sum_\alpha\hat{H}_\alpha(t),
\end{align}
where $\alpha=\text{s,d}$, denotes the source and the drain contacts, respectively.
The tunneling Hamiltonian is given by
\begin{align}\label{eq:tunnelHamiltonian}
\hat{H}_\text{T}=\sum_{\alpha\kappa} t_\alpha\left(\hat{c}^\dag_{\alpha\kappa}\hat{d}+\hat{d}^\dag\hat{c}_{\alpha\kappa}\right),
\end{align}
where $\hat{c}^\dag_{\alpha\kappa}(\hat{c}_{\alpha\kappa})$ creates (annihilates) an electron in lead $\alpha$. The coupling between molecule and leads is parametrized by the tunneling matrix elements $t_\text{s}$ and $t_\text{d}$. Here, we consider the weak coupling regime so that the energy broadening $\hbar{\rm{\Gamma}}$ of molecular levels due to $\hat{H}_\text{T}$ is small, i.e., $\hbar{\rm{\Gamma}}\ll\hbar\omega_0,k_{\text{B}}T$, and a perturbative treatment for $\hat{H}_\text{T}$ in the framework of rate equations is appropriate. For simplicity, we assume that the tunneling amplitude $t_\text{s/d}$ of lead $\text{s/d}$ is real and independent of the momentum $\hbar\kappa$ of the lead state. In addition, we consider a symmetric device with $t_\text{s}=t_\text{d}$.
Finally, the time dependent lead Hamiltonian is described by
\begin{align}\label{eq:leadHamiltonian}
\hat{H}_\alpha(t)=\sum_{\kappa} \bigl[\varepsilon_\kappa +{\rm{\Delta}}\mu_\alpha (t)\bigr]\hat{c}^\dag_{\alpha\kappa}\hat{c}_{\alpha\kappa}.
\end{align}
The above equation describes the lead Hamiltonian of non-interacting electrons with dispersion relation $\varepsilon_\kappa$. The time-varying chemical potential ${\rm{\Delta}}\mu_\alpha (t)$ of lead $\alpha$ depends on the applied bias voltage, and yields a $\kappa$-independent shift of all the single-particle levels.

\subsection{Time dependent master equations for the reduced density matrix}\label{subsec:MasterEquation}

In this section, we briefly derive the equation of motion for the reduced density matrix (RDM) of the molecular junction accounting for the time-dependence, Eq.~\eqref{eq:leadHamiltonian}, of the lead Hamiltonian $\hat{H}_\alpha(t)$. We restrict to the lowest nonvanishing order in the tunneling Hamiltonian. Nevertheless, due to the explicit time dependance in the leads Hamiltonian, this work represents an extension of previous studies on similar systems
(see e.g., Refs.~\cite{84PRB115432,schultz2010,begemannPRB2008,donarininanolett2009,braigPRB2005,braunPRB2004,wunschPRB2005,harbolaPRB2006,mayrhoferEPJB2007,kollerNJP2007,hornberger2008,schultzPRB2009,dominguezPRB2011,brandesPRB2003,blanterPRL2004,husseinPRB2010}). The method is based on the well known Liouville equation for the time evolution of the density matrix of the full system consisting of the leads and the generic quantum dot. To describe the electronic transport through the molecule, we solve the Liouville equation
\begin{align}\label{eq:densitymatrix}
i\hbar\frac{\partial\hat{\rho}_\text{red}^I(t)}{\partial t}=\text{Tr}_\text{leads}\left[{\hat{H}}^I_\text{T}(t),\hat{\rho}^I(t) \right]
\end{align}
for the reduced density matrix $\hat{\rho}_\text{red}(t)=\text{Tr}_\text{leads}\left\{\hat{\rho}(t) \right\}$ in the interaction picture, where the trace over the leads degrees of freedom is taken. In the above equation, $\hat{H}^I_\text{T}(t)$ is the tunneling Hamiltonian in the interaction picture to be calculated as below:
\begin{align}\label{eq:TH}
\hat{H}^I_\text{T}(t)=\sum_{\alpha\kappa}t_\alpha\left[\hat{c}^\dag_{\alpha\kappa}\hat{d}(t)
e^{\frac{i}{\hbar}[\varepsilon_\kappa t+\zeta_\alpha(t)]}+\rm{h.c.}\right],
\end{align}
where $\zeta_\alpha(t)=\int^t_{t_0}{\rm{\Delta}}\mu_\alpha(t'){\rm{d}}t'$.
We make the following approximations to solve the above equation:
(i) The leads are considered as reservoirs of noninteracting electrons in adiabatic thermal equilibrium. Note that this implies that the time scale of variation of the external perturbation has to be large compared to the relaxation time scale of the reservoirs (cf. Eq.~\eqref{eq:A} below). We assume the coupling between system and reservoirs has been switched on at time $t=t_0$ and consider a factorized initial condition. Thus at times $t\ge t_0$ it holds $\rho^I(t)=\rho^I_\text{sys}(t) \otimes \rho_\text{s}\rho_\text{d}+\theta(t-t_0)O(\hat{H}_\text{T}):=\rho^I_\text{sys}(t) \otimes \rho_\text{leads}+\theta(t-t_0)O(\hat{H}_\text{T})$, where the correction in the tunnelling Hamiltonian drops in the second order master equation (see Eq. \eqref{eq:ROH}). Here $\rho_\text{s/d}=\frac{1}{Z_\text{s/d}}e^{-\beta(\hat{H}_\text{s/d}(t)-\mu_\text{s/d}(t)\hat{N}_\text{s/d})}$ denotes the thermal equilibrium grandcanonical distribution of lead $\text{s/d}$, $Z_\text{s/d}$ is the partition function, $\beta$ the inverse of the thermal energy, $\hat{N}_\text{s/d}$ the electron number operator, and $\mu_\text{s/d}(t)=\mu_0+{{\rm{\Delta}}}\mu_\text{s/d}(t)$ is the time dependent chemical potential of lead $\text{s/d}$ which depends on the applied bias voltage.
Note that the levels shift is taken into account by the time-dependent perturbation ${\rm{\Delta}}\mu_\text{s/d}(t)$, while the change in chemical potential is taken into account accordingly via the chemical potential $\mu_\text{s/d}(t)$ so that the net positive or negative charge accumulation in the leads is avoided. Conventionally, we take the molecular energy levels as a fixed reference and let the bias voltage drop across the source and drain contacts through the Fermi energies as~\cite{79PRL2530}
\begin{align}\label{eq:densitymatrix2}
&\mu_\text{s}(t)=\mu_0+(1-\eta)eV_\text{b}(t),\nonumber\\&\mu_\text{d}(t)=\mu_0-\eta eV_\text{b}(t),
\end{align}
where $0\leq\eta\leq 1$ describes the symmetry of the voltage drop across the junction. Specifically, $\eta=0$ corresponds to the most asymmetric situation, while $\eta=1/2$ represents the symmetric case. In addition, we consider a sinusoidally-varying bias voltage, i.e., $V_\text{b}(t)=V_0\sin(\omega t)$, where $\omega$ is the frequency of the driving field.
(ii) Since we assume weak coupling of the molecule to the leads, we treat the effects of $\hat{H}_\text{T}$ perturbatively up to second order. Accounting for the time-evolution as in Eq.~\eqref{eq:TH} of the leads creation/annihilation operators, we find:
\begin{align}\label{eq:ROH}
\dot{\hat{\rho}}^I_\text{red}(t)=&-\sum_{\alpha\kappa}\frac{|t_\alpha|^2}{\hbar^2}\int^t_{t_0} {\rm{d}}t'\biggl\{f_\alpha\left(\varepsilon_\kappa-\mu_0\right)\hat{d}(t)\hat{d}^\dag(t')\hat{\rho}^I_\text{red}(t')
\nonumber\\&\times e^{\frac{i}{\hbar}[\varepsilon_\kappa (t-t')+\zeta_{\alpha}(t)-\zeta_{\alpha}(t')]}+\left[1-f_\alpha\left(\varepsilon_\kappa-\mu_0\right)\right]\nonumber\\&\times \hat{d}^\dag(t)\hat{d}(t')\hat{\rho}^I_\text{red}(t')e^{-\frac{i}{\hbar}[\varepsilon_\kappa (t-t')+\zeta_{\alpha}(t)-\zeta_{\alpha}(t')]}\nonumber\\&
-\left[1-f_\alpha\left(\varepsilon_\kappa-\mu_0\right)\right]\hat{d}(t)\hat{\rho}^I_\text{red}(t')\hat{d}^\dag(t')\nonumber\\&\times e^{\frac{i}{\hbar}[\varepsilon_\kappa(t-t')+\zeta_{\alpha}(t)
-\zeta_{\alpha}(t')]}-f_\alpha\left(\varepsilon_\kappa-\mu_0
\right)\hat{d}^\dag(t)\nonumber\\&\times\hat{\rho}^I_\text{red}(t')\hat{d}(t') e^{-\frac{i}{\hbar}[\varepsilon_\kappa(t-t')+\zeta_{\alpha}(t)-\zeta_{\alpha}(t')]}+\text{h.c}
\biggr\}.
\end{align}
In the derivation of the above equation we have used the relation: $\text{Tr}_\text{leads}\left\{\hat{c}^\dag_{\alpha\kappa}\hat{c}_{\alpha'\kappa'}\hat{\rho}_\text{s}\hat{\rho}_\text{d}\right\}
=\delta_{\alpha\alpha'}\delta_{\kappa\kappa'}f\left(\varepsilon_\kappa-\mu_0\right)$,\\ where $f\left(\varepsilon_\kappa-\mu_0\right)$ is the Fermi function, and the cyclic property of the trace. By summing over $\kappa$ we obtain the generalized master equation (GME) for the reduced density matrix in the form
\begin{align}\label{eq:ROHC}
\dot{\hat{\rho}}^I_\text{red}(t)=&-\sum_{\alpha}\frac{|t_\alpha|^2}{\hbar^2}\int^t_{t_0} {\rm{d}}t'\biggl\{F_\alpha(t-t',\mu_0)\hat{d}(t)\hat{d}^\dag(t')\nonumber\\&\times\hat{\rho}^I_\text{red}(t')
 e^{\frac{i}{\hbar}[\zeta_{\alpha}(t)-\zeta_{\alpha}(t')]}+F_\alpha(t-t',-\mu_0)\nonumber\\&\times \hat{d}^\dag(t)\hat{d}(t')\hat{\rho}^I_\text{red}(t') e^{-\frac{i}{\hbar}[\zeta_{\alpha}(t)-\zeta_{\alpha}(t')]}
\nonumber\\&-F^\ast_\alpha(t-t',-\mu_0)\hat{d}(t)\hat{\rho}^I_\text{red}(t')\hat{d}^\dag(t') \nonumber\\&\times e^{\frac{i}{\hbar}[\zeta_{\alpha}(t)
-\zeta_{\alpha}(t')]}-F^\ast_\alpha(t-t',\mu_0)\hat{d}^\dag(t)\nonumber\\&\times \hat{\rho}^I_\text{red}(t')\hat{d}(t') e^{-\frac{i}{\hbar}[\zeta_{\alpha}(t)-\zeta_{\alpha}(t')]}+\text{h.c}
\biggr\},
\end{align}
where the correlation function $F_\alpha\left(t-t',\mu_0\right)$ of lead $\alpha$ [see Appendix~\ref{app:Correlator}] has, in the wide band limit, the following form:
\begin{align}\label{eq:CF}
F_\alpha\bigl(t-t',\mu_0\bigr)&=\pi\hbar D_\alpha e^{i\frac{\mu_0}{\hbar}(t-t')}\nonumber\\&\times\biggl\{\delta\bigl(t-t'\bigr)-\frac{i}{\hbar\beta\sinh\bigl[\pi\frac{(t-t')}
{\hbar\beta}\bigr]}\biggr\},
\end{align}
which decays with the time difference $t-t'$ approximately as $\exp\left[-\pi\frac{(t-t')}{\hbar\beta}\right]$ on the time scale $\frac{\hbar\beta}{\pi}$. Here $D_\alpha$ is the density of states of lead $\alpha$ at the Fermi level.
(iii) Since we are interested in the long-term dynamical behavior of the system, we set $t_0\to -\infty$ in Eq.~\eqref{eq:ROHC}. Furthermore, we replace $t'$ by $t-t''$. We then apply the Markov approximation, where the time evolution of ${\hat{\rho}}_\text{red}^I$ is taken only local in time, meaning we approximate $\hat{\rho}_\text{red}(t-t'')\sim \hat{\rho}_\text{red}(t)$ in Eq.~\eqref{eq:ROHC}.  In general the condition of time locality requires that~\cite{304PR229}
\begin{align}\label{eq:A}
{\rm{\Gamma}},\omega\ll\frac{\pi}{\hbar\beta}.
\end{align}
Here we defined from Eq.~\eqref{eq:ROHC} together with Eq.~\eqref{eq:CF}, ${\rm{\Gamma}}_\alpha=\frac{2\pi}{\hbar}|t_\alpha|^2D_\alpha$ as the bare transfer rates and $\hbar{\rm{\Gamma}}=\sum_\alpha\hbar{\rm{\Gamma}}_\alpha$ as the tunneling-induced level width. 
Notice that the validity of the Markov approximation, justified in this case, is crucially depending by the order of the current cumulant and the order of the perturbation expansion in the tunnelling coupling \cite{BraggioPRL2006}.
Finally, the condition of adiabatic driving Eq.~\eqref{eq:A} allows to approximate $\zeta_\alpha(t)-\zeta_\alpha(t-t'')={\rm{\Delta}}\mu_\alpha(t)t''$.
Taking into account these simplifications, the generalized master equation (GME) for the reduced density matrix acquires the form
\begin{align}\label{eq:rhomarkov}
\dot{\hat{\rho}}^I_\text{red}(t)&=-\sum_{\alpha}\frac{|t_\alpha|^2}{\hbar^2}\int^\infty_{0} {\rm{d}}t''\biggl\{F[t'',\mu_\alpha(t)]\hat{d}(t)\hat{d}^\dag(t-t'')\nonumber\\&\times\hat{\rho}^I_\text{red}(t)
+F[t'',-\mu_\alpha(t)]\hat{d}^\dag(t)\hat{d}(t-t'')\hat{\rho}^I_\text{red}(t)
\nonumber\\&-F^\ast[t'',-\mu_\alpha(t)]\hat{d}(t)\hat{\rho}^I_\text{red}(t)\hat{d}^\dag(t-t'')\nonumber\\&
-F^\ast[t'',\mu_\alpha(t)]\hat{d}^\dag(t)\hat{\rho}^I_\text{red}(t)\hat{d}(t-t'')+\text{h.c.}\biggr\},
\end{align}
where $F[t'',\mu_\alpha(t)]=F_\alpha(t'',\mu_0)e^{\frac{i}{\hbar}{\rm{\Delta}}\mu_\alpha(t)t''}$.
Since the eigenstates ${|n,m\rangle }_1 $ of $ {\hat{H}}_\text{sys} $ are known, it is convenient to calculate the time evolution of ${\hat{\rho}_\text{red}}^I $ in this basis. For a generic quantum dot system, this projection yields a set of differential equations coupling diagonal (populations) and off-diagonal (coherences) components of the RDM. For the simple Anderson-Holstein model Eq.~\eqref{eq:SysHamiltonian} coherences and populations are, however, decoupled.
In the sequential-tunneling regime, the master equation for the occupation probabilities $P^{m}_n={}_1\langle n,m|\hat{\rho}_\text{red}|n,m\rangle_1$ of finding the system in one of the polaron eigenstates assumes the form 
\begin{align}\label{eq:population}
\dot{P}^{m}_n=& \sum_{n',m'} {{\rm{\Gamma}}}^{m'\to m}_{n'\to n}(t)P^{m'}_{n'}-\sum_{n',m'} {\rm{\Gamma}}^{m\to {m'}}_{n\to {n'}}(t)P^{m}_n,
\end{align} 
where the inequality $\Gamma \ll \omega_0$ ensures the applicability of the secular approximation, i.e., the separation between the dynamics of populations and coherences. In the numerical treatment of these equations we truncate the phonon space. Convergence is reached already with 40 excitations.
 In Eq. \eqref{eq:population} the coefficient ${\rm{\Gamma}}^{m'\to m}_{n'\to n}$ denotes the transition rate from $|n',m'\rangle_1$ into the many body state $|n,m\rangle_1$, while ${\rm{\Gamma}}^{m\to {m'}}_{n\to {n'}}$ describes the transition rate out of the state $|n,m\rangle_1$ to $|n',m'\rangle_1$.  Taking into account all possible single-electron-tunneling processes, we obtain the incoming and outgoing tunneling rates, in the wide band limit, as
\begin{align}\label{eq:incomingrate}
{\rm{\Gamma}}^{m\to {m'}}_{0\to 1}(t)&=\sum_\alpha{\rm{\Gamma}}_\alpha F_{mm'}f^+\bigl[\varepsilon+\hbar\omega_0\left(m'-m\right)-\mu_\alpha(t)\bigr]\nonumber\\&\ \equiv\sum_\alpha {\rm{\Gamma}}^{m\to {m'}}_{\alpha,0\to 1}(t),
\end{align}
\begin{align}\label{eq:outgoingrate}
{\rm{\Gamma}}^{m'\to m}_{1\to 0}(t)&=\sum_\alpha{\rm{\Gamma}}_\alpha F_{mm'}f^-\bigl[\varepsilon+\hbar\omega_0\left(m'-m\right)-\mu_\alpha(t)\bigr]\nonumber\\&\ \equiv\sum_\alpha {\rm{\Gamma}}^{m'\to {m}}_{\alpha,1\to 0}(t),
\end{align}
where the terms describing sequential tunneling from and to the lead $\alpha$ are proportional to the Fermi functions $f^+(x-\mu_\alpha)=f(x-\mu_\alpha)$ and $f^-(x-\mu_\alpha)=1-f(x-\mu_\alpha)$, respectively.
Notice that the integrations over energy and time introduce the explicit time dependance in the Fermi functions.
The factor $F_{mm'}=|\langle m|\hat{X}|m'\rangle|^2$ is the Franck-Condon matrix element which can be calculated, with $\hat{X}$ defined in Section~\ref{sec:PolaronTransformation}, explicitly using Appendix~\ref{app:Transitionrates}. The sum rules $\sum_m F_{mm'}=\sum_{m'}F_{mm'}=1$ are well satisfied because of the completeness of each vibrational basis set ${|0,m}\rangle$ and ${|1,m'}\rangle_1$. This factor describes the wave-function overlap between the vibronic states participating in the particular transition. It contains essential information about the quantum mechanics of the molecule and significantly influences the transport properties of the single-molecule junction.
Within the rate-equation approach, the (particle) current through lead $\alpha$ is determined by
\begin{align}\label{eq:C}
I_\alpha(t) = & \sum_{mm'} \left({\rm{\Gamma}}^{m\to {m'}}_{\alpha,0\to 1}(t) P^{m}_{0}(t)-{\rm{\Gamma}}^{m'\to m}_{\alpha,1\to 0}(t) P^{m'}_{1}(t) \right)
\end{align}
and it is in general time dependent. Moreover, differently from the stationary case, in general $I_L(t) \neq -I_R(t)$. The charge is though not accumulating on the dot since, for the average quantities

\begin{equation}
I_{\alpha,\rm av} = \lim_{t \to \infty}\int_{t}^{t+T_{\rm ex}} {\rm d}t' I_{\alpha}(t')
\end{equation}
 it holds $I_{L,{\rm av}} = -I_{R, {\rm av}}$, as it can be easily proved considering that the average charge on the dot oscillates with the same period $T_{\rm ex}$ of the driving bias. Finally, in the DC limit $\omega \to 0$ the relation $I_{L}(t) = -I_{R}(t)$ holds as the fully adiabatic driving allows to reach the quasi-stationary limit at all times.

\section{Lifetimes and bistability of states }\label{sec:Lifetimes}

\begin{figure}[ht]
\centerline{\psfig{figure=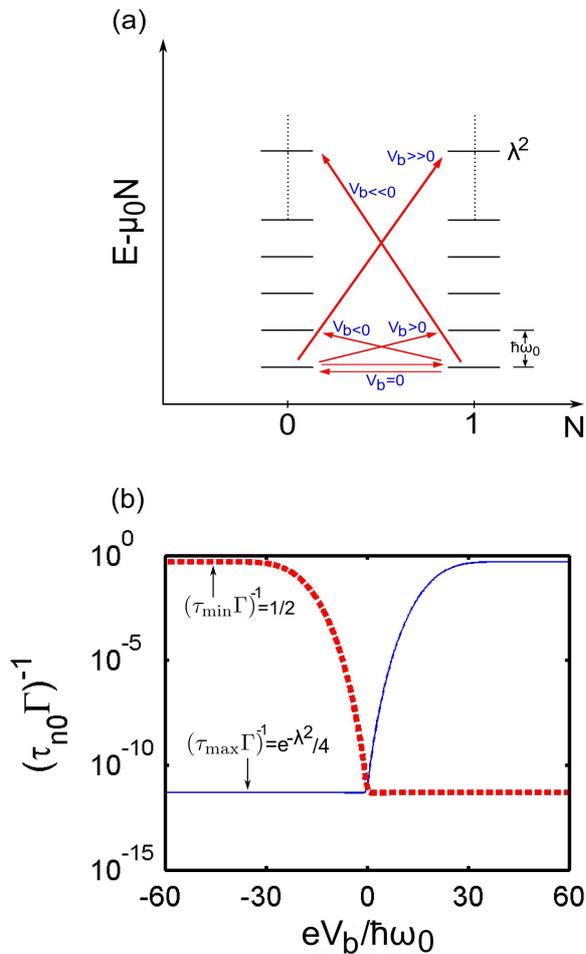,width=8cm}}
\caption{(Color online). (a) Energy-level scheme for the relevant transitions in a coordinate system given by the particle number $N$ and grandcanonical energy $E-\mu_0N$ at $V_\text{b}=0$. The red lines represent the transitions threshold, where the thickness of each transition line gives the strength of the transition. The polaron energy levels are aligned with the 0-particle states for our chosen set of parameters ($\mu_0=0,\ V_\text{g}=0,\ \varepsilon_0=25\hbar\omega_0,\ \lambda=5$) yielding the polaron shift $\varepsilon_\text{p}=\varepsilon_0$. (b) Inverse lifetimes $\left(\tau_{n0}{\rm{\Gamma}}\right)^{-1}$ on logarithmic scale as a function of normalized bias voltage $eV_\text{b}/\hbar\omega_0$. The red thick line represents the inverse lifetime of the 1-particle ground state, while the thin blue line refers to the 0-particle ground state.}
\label{fig:TS}
\end{figure}

In this section, we show that when the bias voltage drop is asymmetric across the junction, upon sweeping the bias, one can tune the lifetime of the neutral and charged states to achieve a bistable system.
The lifetime of a state is obtained by calculating the switching rate of that state.
The lifetime $\tau_{nm}$ of a generic quantum state $|n,m\rangle_1$ is given by the sum of the rates of all possible processes which depopulate this state, i.e.,
\begin{align}\label{eq:lifetime}
\tau^{-1}_{nm}=\sum_{n',m'}{\rm{\Gamma}}^{m\to {m'}}_{n\to {n'}},
\end{align}
and it defines, at least on a relative scale, the stability of the state $|n,m\rangle_1$.
Thus, at finite bias voltage, the inverse lifetime of the 0-particle $m$th vibronic state is given by the relation
\begin{align}\label{eq:tau0m}
\tau^{-1}_{0m}=\sum_{\alpha,m'}{\rm{\Gamma}}_\alpha F_{mm'}f^+\bigl[\varepsilon+\hbar\omega_0\left(m'-m\right)-\mu_\alpha\bigr].
\end{align}
In a similar way, the inverse lifetime of the 1-particle and $m$th vibronic state is expressed as
\begin{align}\label{eq:tau1mp}
\tau^{-1}_{1m}=\sum_{\alpha,m'}{\rm{\Gamma}}_\alpha F_{mm'}f^-\bigl[\varepsilon+\hbar\omega_0\left(m-m'\right)-\mu_\alpha\bigr].
\end{align}
A consequence of Eqs.~\eqref{eq:tau0m} and~\eqref{eq:tau1mp} is that, due to the characteristic features of the Franck-Condon matrix elements, in the strong electron-vibron coupling regime, the tunneling with small changes in $m-m'$ is suppressed exponentially. Hence only some selected vibronic states contribute to the tunneling process. However, tunneling also depends on the bias voltage and temperature through the Fermi function.
To proceed further, let us focus first on the lifetime of the 0- and 1-particle ground states for the case of fully asymmetric coupling of the bias voltage to the leads, i.e., $\eta=0$:
\begin{align}\label{eq:tau00Asyground}
\tau^{-1}_{00}=&\sum_{m'}\frac{e^{-\lambda^2}\lambda^{2m'}}{m'!}\bigl\{{\rm{\Gamma}}_\text{s} f^+\left(\varepsilon+m'\hbar\omega_0-\mu_0-eV_\text{b}\right)\nonumber\\&+{\rm{\Gamma}}_\text{d} f^+\left(\varepsilon+m'\hbar\omega_0-\mu_0\right)\bigr\},
\end{align}
\begin{align}\label{eq:tau10Asyground}
\tau^{-1}_{10}=&\sum_{m'}\frac{e^{-\lambda^2}\lambda^{2m'}}{m'!}\bigl\{{\rm{\Gamma}}_\text{s} f^-\left(\varepsilon-m'\hbar\omega_0-\mu_0-eV_\text{b}\right)\nonumber\\&+{\rm{\Gamma}}_\text{d} f^-\left(\varepsilon-m'\hbar\omega_0-\mu_0\right)\bigr\}.
\end{align}
One can see from Eq.~\eqref{eq:tau00Asyground} that if in the considered parameters range is $\varepsilon+m'\hbar\omega_0\gg \mu_0$, i.e., $f\left(\varepsilon+m'\hbar\omega_0-\mu_0\right)\rightarrow 0$, then the second term in the bracket is negligible. The first term is nonzero at large positive bias, while at large negative bias it remains negligible. In a similar way one can analyze the behavior of $\tau_{10}^{-1}$ in which the first term on the r.h.s. of Eq.~\eqref{eq:tau10Asyground} will be dominating at large negative bias. In order to understand the mechanism of this process the energy-level scheme for the relevant transitions in a coordinate system given by the particle number $N$ and the grandcanonical energy $E-\mu_0N$ shown in Figure~\ref{fig:TS}. We choose $V_\text{g}=0$ and $\mu_0=0$. Moreover, the polaron energy levels are at resonance with the 0-particle states for our chosen set of parameters: we set $\varepsilon_\text{p}=\varepsilon_0$ and hence $\varepsilon=0$. Then the only transitions allowed at zero bias are ground state $\leftrightarrow$ ground state transitions. At finite bias also transitions involving excited vibronic states become allowed. In particular, at $V_\text{b}=0$ it follows from Eqs.~\eqref{eq:tau00Asyground}, and~\eqref{eq:tau10Asyground} that

\begin{align}\label{eq:STC1}
\tau^{-1}_{00}(V_\text{b}=0)=\tau^{-1}_{10}(V_\text{b}=0)=e^{-\lambda^2}({\rm{\Gamma}}_\text{s}+{\rm{\Gamma}}_\text{d})/2,
\end{align}
while at $|V_\text{b}|\to\infty$ it holds
\begin{align}\label{eq:STC2}
\tau^{-1}_{00}(V_\text{b}\to\infty)&=\tau^{-1}_{10}(V_\text{b}\to-\infty)\nonumber\\&={\rm{\Gamma}}_\text{s}+\frac{{\rm{\Gamma}}_\text{d}}{2}
e^{-\lambda^2}\sim{\rm{\Gamma}}_\text{s}\equiv\tau^{-1}_\text{min},
\end{align}
whereas
\begin{align}\label{eq:STC3}
\tau^{-1}_{00}(V_\text{b}\to-\infty)=\tau^{-1}_{10}(V_\text{b}\to\infty)=\frac{{\rm{\Gamma}}_\text{d}}{2}e^{-\lambda^2}\equiv\tau^{-1}_\text{max}.
\end{align}
In practice the asymptotic behaviors are already reached at $e|V_\text{b}|/\hbar\omega_0\sim 2\lambda^2$ as observed in Figure~\ref{fig:TS}(b). Note that $\tau_\text{max}$ and $\tau_\text{min}$ set the maximum and minimum achievable lifetimes which, due to $\tau_\text{max}/\tau_\text{min}\sim e^{\lambda^2}$, can differ by several orders of magnitude for $\lambda>1$. Note also that near zero bias the lifetimes are so long that the system never likes to charge or discharge and a bistable situation is reached. A selective switching, however, can occur upon sweeping the bias voltage. Hence $\tau_\text{min}$ also sets the time scale for switching: $\tau_\text{min}\sim\tau_\text{switch}$.

\begin{figure}[ht]
\centerline{\psfig{figure=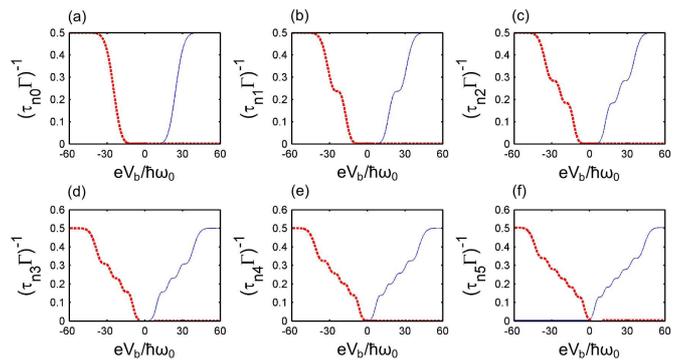,width=\columnwidth}}
\caption{(Color online). Inverse lifetime $\left(\tau_{nm}{\rm{\Gamma}}\right)^{-1}$ as a function of normalized bias voltage $eV_\text{b}/\hbar\omega_0$ for (a) vibronic ground states, (b) first excited states, (c) second excited states, (d) third excited states, (e) fourth excited states, (f) fifth excited states when $V_\text{g}=0$. The blue thin line represents the inverse lifetime of the 0-particle state ($n=0$), while the thick dashed red line refers to the 1-particle state ($n=1$). The asymmetry parameter is $\eta=0$ and we fix the zero of the energy at the leads chemical potential at zero bias: $\mu_0=0$. The energy of the molecular level is $\varepsilon_0=25\hbar\omega_0$. The electron-vibron coupling constant is $\lambda=5$ yielding a polaron shift $\varepsilon_\text{p}=\varepsilon_0$. Finally, the thermal energy is $k_\text{B}T=0.2\hbar\omega_0$, the frequency of the driving field is $\omega=0.002\omega_0$, and ${\rm{\Gamma}}_\text{s}={\rm{\Gamma}}_\text{d}=0.006\omega_0$.}
\label{fig_s_t}
\end{figure}

\begin{figure}[ht]
\centerline{\psfig{figure=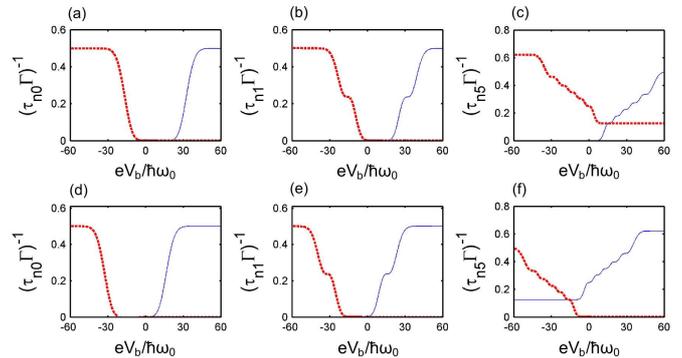,width=\columnwidth}}
\caption{(Color online). Inverse lifetime $\left(\tau_{nm}{\rm{\Gamma}}\right)^{-1}$ as a function of normalized bias voltage $eV_\text{b}/\hbar\omega_0$ for (a) vibronic ground states, (b) first excited states, (c) fifth excited states when $eV_\text{g}/\hbar\omega_0=8$, while (d) shows vibronic ground states, (e) first excited states, and (f) fifth excited states, when $eV_\text{g}/\hbar\omega_0=-8$. The remaining parameters are the same as used in Figure~\ref{fig_s_t}.}
\label{fig_s_tb}
\end{figure}
Analogously, we can explain the behavior of the lifetimes of the excited states (see Figure~\ref{fig_s_t}).
It follows that in the considered parameters range, in general, the 0-particle vibronic states are stable at large enough negative bias voltage, while the 1-particle vibronic states are stable at large positive bias. There is, however, an interval of bias voltage, the so-called bistable region, where both states $|1,m'\rangle_1$ and $|0,m\rangle_1$ are metastable for not too large $m$ and $m'$, as shown in Figure~\ref{fig_s_t}.
Moreover, $m$ steps are observed in the inverse lifetimes $\tau^{-1}_{nm}$ (see Figures~\ref{fig_s_t}(b-f)) because for certain values of the coupling constant $\lambda$ some of the FC factors $F_{mm'}$ vanish or are exponentially small such that the additional channels opened upon increasing the bias voltage do not have pronounced contribution. For instance, the FC factor for the first excited vibronic state can be described as
\begin{align}\label{eq:F1}
F_{1m'}=e^{-\lambda^2}\frac{\lambda^{2(m'-1)}}{m'!}(m'-\lambda^2)^2,
\end{align}
which vanishes for $m'=\lambda^2$. That is why a plateau around $eV_\text{b}/\hbar\omega_0=25$ in Figure~\ref{fig_s_t}(b) is observed for our chosen parameters. Analogously, using Eq.~\eqref{eq:F2m}, one can find (cf. Appendix~\ref{app:F2m}) that $F_{2m'}$ has two minima at
\begin{align}\label{eq:F2}
m_1=\frac{1+2\lambda^2+\sqrt{1+4\lambda^2}}{2},\ m_2=\frac{1+2\lambda^2-\sqrt{1+4\lambda^2}}{2}.
\end{align}
Hence two plateau can be observed (see Figure~\ref{fig_s_t}(c)) around $eV_\text{b}/\hbar\omega_0=20$ and $eV_\text{b}/\hbar\omega_0=31$.
Similar arguments can be extended to explain the steps in the inverse lifetimes of higher excited states.
This also implies that the bias window for bistability shrinks for excited states and even disappears for large enough $m$. It follows that the major contribution in bistability is coming from low excited vibronic states.
Note that the bistability of the many body states is crucial for the hysteresis and hence memory effects which is discussed in the next section.
Finally, a closer inspection of Figure~\ref{fig_s_t} reveals that the minimum of the inverse lifetime increases with the vibronic quantum number $m$. This effect can be understood easily by analyzing the minimum of the inverse lifetime for each particle state. For example the minimum of the inverse lifetime for the 0-particle vibronic ground state is, cf Eq.~\eqref{eq:STC3}, whereas for the 0-particle vibronic first excited state is
\begin{align}\label{eq:MG1}
\tau^{-1}_{01}(V_\text{b}\to-\infty)=\frac{\rm{\Gamma}_\text{d}}{2}(1+\lambda^4)e^{-\lambda^2}.
\end{align}
From Eqs.~\eqref{eq:STC3} and~\eqref{eq:MG1}, one can conclude that $\tau^{-1}_{00}(V_\text{b}\to-\infty)<\tau^{-1}_{01}(V_\text{b}\to-\infty)$.
A similar explanation can be extended to the higher excited states.
For gate voltages such that $eV_\text{g}>0$, the 1-particle vibronic excited states are becoming unstable faster than the 0-particle states (see Figure~\ref{fig_s_tb}(a)-(c)), while for large negative gate ($eV_\text{g}<0$), the 0-particle states are getting unstable fast (see Figure~\ref{fig_s_tb}(d)-(f)).
In order to explain this effect, we analyze the shift of the inverse lifetime of the 0-particle vibronic first excited state, $\tau^{-1}_{01}$, as follows:\\
The maximum of the inverse lifetime for $V_\text{g}\neq 0$ is
\begin{align}\label{eq:MGS1}
\tau^{-1}_{01}(V_\text{b}\to\infty)={\rm{\Gamma}}_\text{s}+{\rm{\Gamma}}_\text{d}\sum_mF_{1m}f(eV_\text{g}+\hbar\omega_0(m-1)),
\end{align}
whereas the minimum is given by
\begin{align}\label{eq:MGS2}
\tau^{-1}_{01}(V_\text{b}\to-\infty)={\rm{\Gamma}}_\text{d}\sum_mF_{1m}f(eV_\text{g}+\hbar\omega_0(m-1)).
\end{align}
Eqs.~\eqref{eq:MGS1} and~\eqref{eq:MGS2} imply that both minimum and maximum of $\tau^{-1}_{01}$ shift by an equal amount and the condition of the bistability region can be tuned by setting $V_\text{g}$.
\section{Quantum switching and hysteresis}\label{sec:SwitchingHysteresis}

Neutral and charged (polaron) states correspond to different potential energy surfaces and transitions between low-lying vibronic states are strongly suppressed in the presence of strong electron-vibron interaction. This leads to the bistability  of the system. Upon applying an external voltage, one can change the state of this bistable system obtaining under specific conditions  hysteretic charge-voltage and current-voltage curves. Here it is crucial to point out that only if the time scale of variation of the external perturbation is shorter than the maximum lifetime but longer than the minimum lifetime of the system hysteresis can be observed, i.e., $\tau_\text{min}\sim\tau_\text{switch}<T_\text{ex}<\tau_\text{max}$.
Due to $\tau_\text{max}>T_\text{ex}$, the system stays in the stable state during the sweeping until the sign of the perturbation changes, the former stable state becomes unstable and, due to $T_\text{ex}<\tau_\text{min}$, a switching to the new stable state can occur. In this section we now consider the situation when $\omega\sim\rm{\Gamma}$, i.e., $T_\text{ex}\sim\tau_\text{switch}$ while in Section~\ref{sec:LFC} the regime $\omega\ll\rm{\Gamma}$, i.e., $T_\text{ex}\gg\tau_\text{switch}$ is addressed.\\

\begin{figure}[h]
\centerline{\psfig{figure=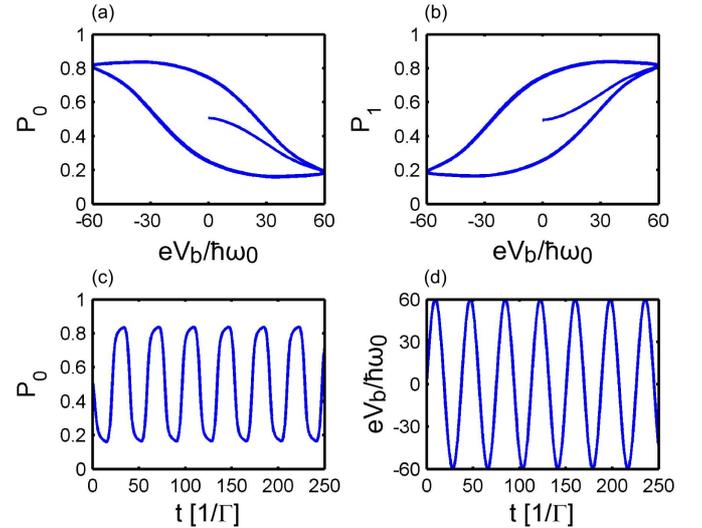,width=\columnwidth}}
\caption{(Color online). (a)-(b) Occupation probabilities $P_0$ and $P_1$ of the 0- and 1-particle electronic states as a function of normalized time dependent bias voltage $eV_\text{b}/\hbar\omega_0$, (c) population of the 0-particle configuration as a function of time; (d) normalized bias voltage as a function of time. The parameters are the same as used in Figure~\ref{fig_s_t}.}
\label{fig_p_t}
\end{figure}

\begin{figure}[h]
\centerline{\psfig{figure=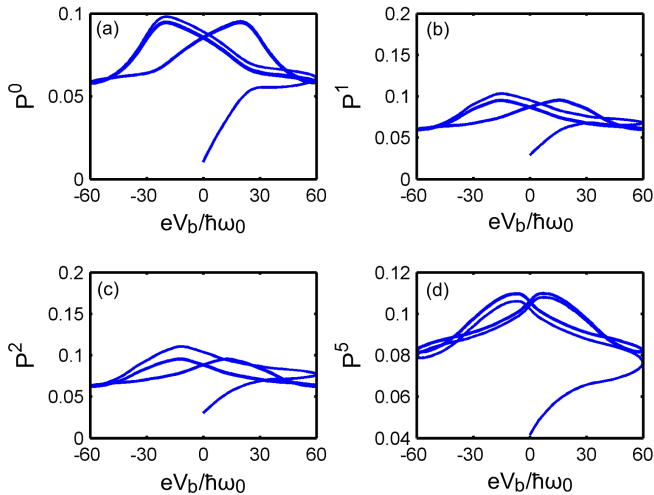,width=\columnwidth}}
\caption{(Color online). Populations $P^m$ of the vibronic states as a function of normalized time dependent bias voltage $eV_\text{b}/\hbar\omega_0$ for (a) ground state, (b) first excited state, (c) second excited state, and (d) fifth excited state. The parameters are the same as used in Figure~\ref{fig_s_t}.}
\label{fig_v_t}
\end{figure}

In Figures \ref{fig_p_t} and \ref{fig_v_t} we present the populations of the electronic states, $P_n=\sum_m P^m_n$, as well as of the vibronic states, $P^m=\sum_n P^m_n$, respectively. Specifically, in Figure~\ref{fig_p_t}(a)-(b), we have plotted the populations of the 0- and 1-particle electronic states as a function of normalized bias voltage, where hysteresis loops can be seen. In Figure~\ref{fig_p_t}(c), instead, we have shown the population of the 0-particle electronic state as a function of time. The latter can be used to determine the time $\tau_\text{switch}$ of switching between the neutral and charged states. In a similar way, the sweeping time $T_\text{ex}$ of the bias voltage can be calculated using Figure~\ref{fig_p_t}(d). By comparison of these two time scales, it is apparent that the switching time is of the same order as the sweeping time and much shorter than the lifetime in the bistable region (see Figure~\ref{fig:TS}). The relation $\tau_\text{switch} \approx T_\text{ex}$ also explains why the switching between the neutral and charged state is on average never complete ($P_0$ oscillates between $0.2$ and $0.8$).

\begin{figure}[h]
\centerline{\psfig{figure=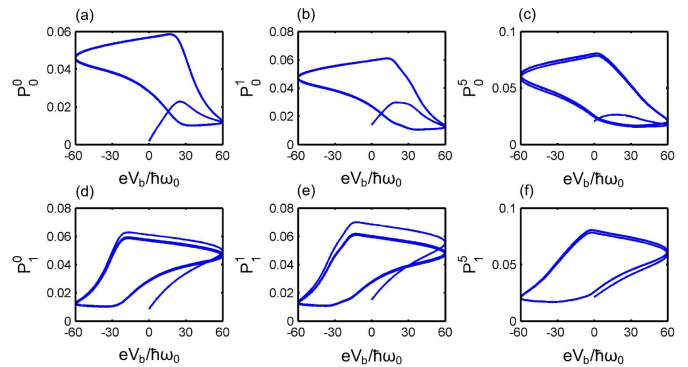,width=\columnwidth}}
\caption{(Color online). Plots of the population $P^m_n$ as a function of normalized time dependent voltage $eV_\text{b}/\hbar\omega_0$ for the 0-particle vibronic (a) ground state, (b) first excited state, (c) fifth excited state, and for the 1-particle vibronic  (d) ground state, (e) first excited state, (f) fifth excited state. The parameters are the same as used in Figure~\ref{fig_s_t}.}
\label{fig_ev_t}
\end{figure}

In Figure~\ref{fig_v_t}, the populations of the vibronic states as a function of the normalized bias voltage are shown, while in Figure~\ref{fig_ev_t} the populations of the different vibronic states resolved for different charges have been plotted. Clearly not only the vibronic ground states (which were considered in Ref.~\cite{78PRB085409}) show hysteretic behavior but the vibronic excited states also exhibit these interesting features. Furthermore, inspection of these figures reveals that even after relaxation on the stable limit cycle, the vibronic excited states are highly populated in the non-stationary case in contrast to the stationary case $\omega \to 0$ (see e.g., Figures~\ref{fig_VP_S} and \ref{fig_EVP_S}) where the population of the excited states is strongly suppressed. Finally, while the general trend is a reduction of the population, the higher the excitation and the populations are negligible for $m \approx 40$, an interesting behaviour can be recognized in the form of the limit cycles. Namely, upon sweeping the bias we find that, for $m \gg 8$ the probability grows at large biases, it stays  essentially constant for $m \approx 8$ and it decreases at larger biases for $m < 8$. The interpretation of this behaviour is still unclear to us. All these observation confirm, though, that it is natural to take into account the vibronic excited states in the dynamics of the system.

\subsection{$I-V$ characteristics}\label{subsec:Current}

\begin{figure}[h]
\centerline{\psfig{figure=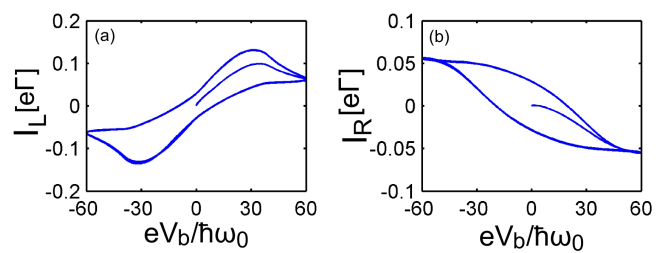,width=\columnwidth}}
\caption{(Color online). Time dependent current as a function of normalized voltage for (a) left lead, (b) right lead. The parameters are the same as used in Figure~\ref{fig_s_t}.}
\label{fig_I_t}
\end{figure}

The hysteretic behavior of the bistable system is also reflected in the current as a function of normalized bias (see Figure~\ref{fig_I_t}) where a hysteresis loop (single loop) is observed in the current calculated both at the left and the right lead. Interestingly, the left and the right currents differ by more than a sign, in contrast to the stationary case. This behavior is understandable again in terms of relaxation time scales. In fact, for voltages  $|V_b|$ outside the bistable region the system relaxes to the stationary regime on a time scale $\tau_{\rm switch}$. Though, since the driving time $T_{\rm ex}$ has the same order of magnitude, the stationary regime cannot be reached. Yet, no net charge accumulation occurs since $I_{L, {\rm av}} = -I_{ R, {\rm av}}$.

\begin{figure}[h]
\centerline{\psfig{figure=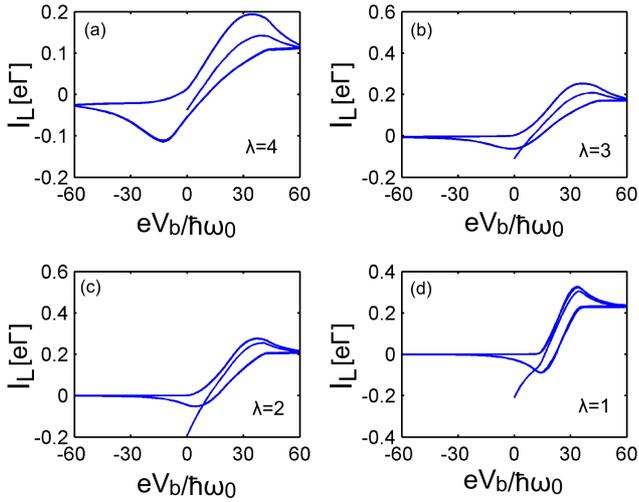,width=\columnwidth}}
\caption{(Color online). Time dependent current for the left lead as a function of the normalized voltage for coupling constants (a) $\lambda=4$, (b) $\lambda=3$, (c) $\lambda=2$, and (d) $\lambda=1$. The remaining parameters are the same as used in Figure~\ref{fig_s_t}.}
\label{fig_I_lam}
\end{figure}

In Figure~\ref{fig_I_lam}, we plot the left time dependent current as a function of the normalized bias for different values of the electron-vibron coupling constant. An inspection of this figure reveals that the width of the hysteresis loop decreases and shifts from zero bias upon decreasing the coupling constant $\lambda$. This feature can be understood by observing that for $\lambda\neq 5$ the polaron shift $\varepsilon_\text{p}$ does not longer compensate the energy of the molecular level $\varepsilon_0$, and hence the polaron energy $\varepsilon\neq 0$.
In other words, the system is no longer behaving symmetrically upon exchange of the sign of the bias voltage. If we consider e.g. the case $\lambda=1$ is, for $V_\text{g}=\mu_0=0,\ \varepsilon/\hbar\omega_0=24$. In turn this implies that $\tau^{-1}_{00}(V_\text{b}=0)\sim 0$ and $\tau^{-1}_{10}(V_\text{b}=0)\sim {\rm{\Gamma}}_\text{s}+{\rm{\Gamma}}_\text{d}$, i.e., the region around zero bias is no longer bistable as for the case $\lambda=5$. Hence the dot is preferably empty at zero bias. Switching however can be reached upon increasing $V_\text{b}$ in the region around $eV_\text{b}\sim\varepsilon$. Overall however the bistability region has shrunk. Similar considerations apply to the other considered values of $\lambda$.

\subsection{Vibron energy}\label{subsec:VibronEnergy}

\begin{figure}[h]
\centerline{\psfig{figure=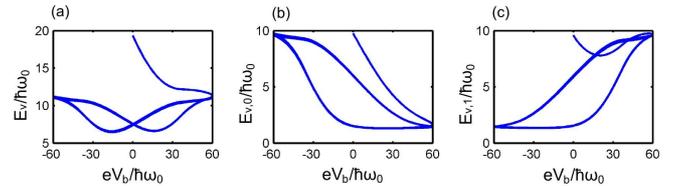,width=\columnwidth}}
\caption{(Color online).  (a) Total vibron energy as a function of the time dependent bias voltage. (b) Vibron energy for the 0-particle, and (c) for the 1-particle configuration only. Parameters are the same as used in Figure~\ref{fig_s_t}.}
\label{fig_v_e}
\end{figure}

In this section, we illustrate the role played by the vibronic energy in the hysteretic behavior of the system. The vibron energy of the whole system can be expressed as
\begin{align}\label{vibronenergy}
E_\text{v}=\text{Tr}_\text{sys}\left\{\hat{\rho}_\text{red}\hbar\omega_0\left(\hat{a}^\dag\hat{a}+\frac{1}{2}\right)\right\},
\end{align}
where the trace is taken over the system degrees of freedom.
The normalized vibronic energy as a function of normalized bias voltage is depicted in Figure~\ref{fig_v_e}(a), where hysteretic loops are also observed. The value of the vibronic energy, together with the observation that the probability distribution is relatively flat over the excitations (see Fig.~\ref{fig_ev_t}) ensures that, depending on the bias, between $10$ and $20$ vibronic excited states are considerably populated. Further insight in the dynamics of the system is obtained by considering the correlation between the vibronic energy and the charge occupation.

The vibron energy associated with the 0-particle state is determined by the relation
\begin{align}\label{vibronenergy0}
E_\text{v,0}=\text{Tr}_\text{sys}\left\{\hat{\rho}_0\hbar\omega_0\left(\hat{a}^\dag\hat{a}+\frac{1}{2}\right)\right\},
\end{align}
with $\hat{\rho}_0=\hat{\rho}_\text{red}|0,m\rangle_1 {}_1\langle 0,m|$.
In Figure~\ref{fig_v_e}(b), the normalized vibronic energy as a function of normalized bias voltage for the 0-particle configuration has been plotted. The hysteresis loop resembles that of Figure~\ref{fig_p_t}(a) implying a direct correlation between the vibronic energy and the population of the neutral state i.e., the more the neutral state is occupied the higher is the associated vibronic energy. Qualitatively the result can be explained as follows: transitions from the charged to the neutral states are predominantly involving low energy charged states and highly excited neutral states. Due to energy conservation and asymmetric bias drop these transitions are confined to the large negative biases where the highly excited neutral states show also a long life time. This situation remains roughly unchanged during the up sweep of the bias until the symmetric condition is obtained at high positive bias and the charged excited states are maximally populated. Finally, the bistability around zero bias explains the hysteresis.
The analytical expression for the vibronic energy of the 1-particle state is given by
\begin{align}\label{vibronenergy1}
E_\text{v,1}=\text{Tr}_\text{sys}\left\{\hat{\rho}_1\hbar\omega_0\left(\hat{a}^\dag\hat{a}+\frac{1}{2}\right)\right\},
\end{align}
with $\hat{\rho}_1=\hat{\rho}_\text{red}|1,m\rangle_1 {}_1\langle 1,m|$. The normalized average vibron energy as a function of normalized bias voltage for the 1-particle configuration is sketched in Figure~\ref{fig_v_e}(c), where we can observe a hysteresis loop resembling that of Figure~\ref{fig_p_t}(b).

In conclusion, the vibron energies also show hysteretic behavior, in analogy to the population-voltage and current-voltage curves, in the non-stationary limit.

\section{Testing lower driving frequencies}\label{sec:LFC}
\begin{figure}[h]
\centerline{\psfig{figure=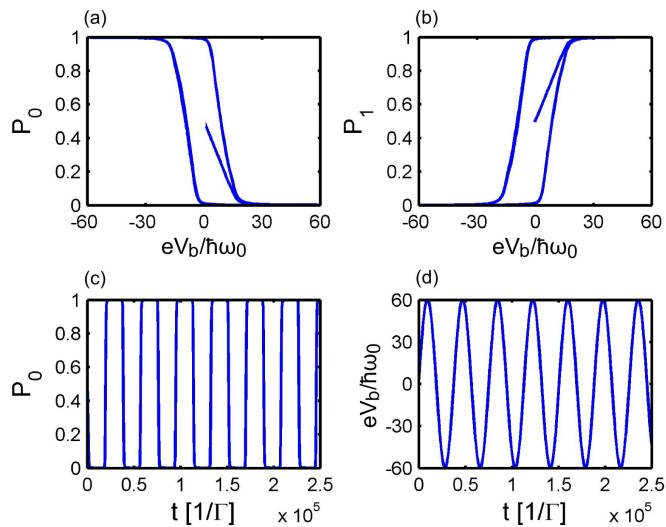,width=\columnwidth}}
\caption{(Color online). (a)-(b) Population of the 0- and 1-particle electronic states as a function of the bias voltage, (c) population of the 0-particle electronic state as a function of time, and (d) normalized bias voltage as a function of time. The frequency of the driving field is $\omega=2\times 10^{-6}\omega_0$. The other parameters are the same as used in Figure~\ref{fig_s_t}.}
\label{fig_p_l}
\end{figure}

\begin{figure}[h]
\centerline{\psfig{figure=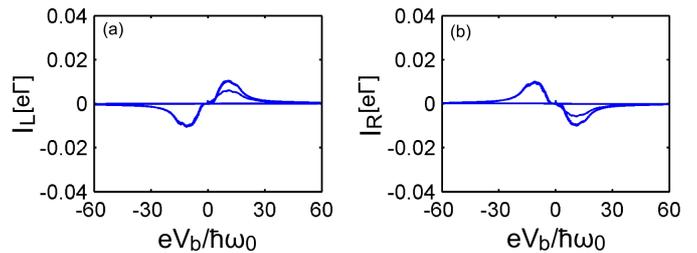,width=\columnwidth}}
\caption{(Color online). (a) Plot of the time dependent current as a function of normalized voltage, (b) current averaged over one driving period as a function of normalized voltage. The value of gate voltage is $V_\text{g}=0$ and the frequency of the driving field is $\omega=2\times 10^{-6}\omega_0$. The other parameters are the same as used in Figure~\ref{fig_s_t}.}
\label{fig_c_l}
\end{figure}

\begin{figure}[h]
\centerline{\psfig{figure=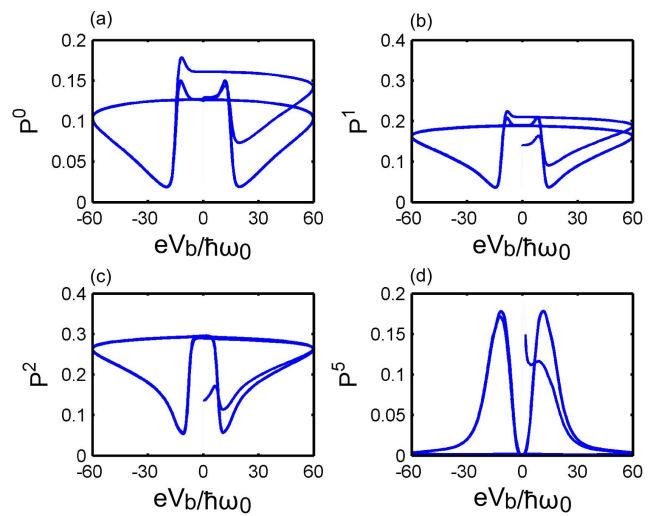,width=\columnwidth}}
\caption{(Color online). Plots of populations of vibronic (a) ground state, (b) first excited state, (c) second excited state, and (d) fifth excited state. The frequency of the driving field is $\omega=2\times 10^{-6}\omega_0$. The other parameters are the same as used in Figure~\ref{fig_s_t}.}
\label{fig_v_l}
\end{figure}

When lowering the driving frequency $\omega$ ($\omega\ll\rm{\Gamma}$) of the external perturbation, we choose $\omega= 2\times 10^{-6}\omega_0$, our model displays features similar to those presented in Ref.~\cite{78PRB085409}. In more detail, we show the population of the electronic states as a function of normalized bias and time in Figure~\ref{fig_p_l}(a)-(b), Figure~\ref{fig_p_l}(c), respectively, whereas in Figure~\ref{fig_p_l}(d) the normalized bias as a function of time is shown. In this case the population-voltage curve is slightly different from Figure~\ref{fig_p_t} because the transition between 0 and 1 occurs more abruptly as a function of $V_\text{b}$ and it is complete. Indeed, for the parameter chosen in Figure~\ref{fig_p_l} is $\varepsilon=0$ and $\tau^{-1}_\text{max}\sim \omega_\text{min}\ll\omega\ll\rm{\Gamma}\sim\tau^{-1}_\text{switch}$. In other words the frequency is small compared to the charge/discharge rate. The system thus follows adiabatically the changes of the bias voltage and only switches at those values of the bias where $\tau_{n0}\sim\tau_\text{switch}$.
The time-dependent left current as a function of normalized bias is shown in Figure~\ref{fig_c_l}(a) giving two loops, one for positive bias sweeping and the other for negative sweeping. The right current is shown in Figure~\ref{fig_c_l}(b). Due to the extremely low frequency the currents substantially fulfill the quasi-stationary relation $I_L(t) = -I_R(t)$ associated to a fully adiabatic regime.
In Figure~\ref{fig_v_l}, we present the populations of the vibronic states and hysteretic loops are visible.
Vibronic states with quantum numbers up to $\lambda$ all display nonvanishing populations, much less than in the case $T_{\rm ex} \approx \tau_{\rm switch}$.
\section{The DC-case ($\omega \to 0$)}\label{sec:stationary}
\begin{figure}[h]
\centerline{\psfig{figure=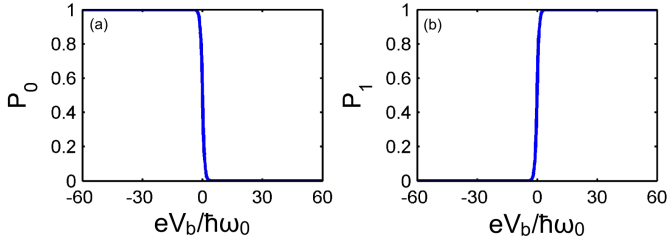,width=\columnwidth}}
\caption{(Color online). Population of (a) the 0-particle electronic state, (b) the 1-particle electronic state. The value of gate voltage is $V_\text{g}=0$, and the frequency of the driving field is $\omega \ll 1/\tau_{\rm max}$. The other parameters are the same as used in Figure~\ref{fig_s_t}.}
\label{fig_P_S}
\end{figure}

\begin{figure}[h]
\centerline{\psfig{figure=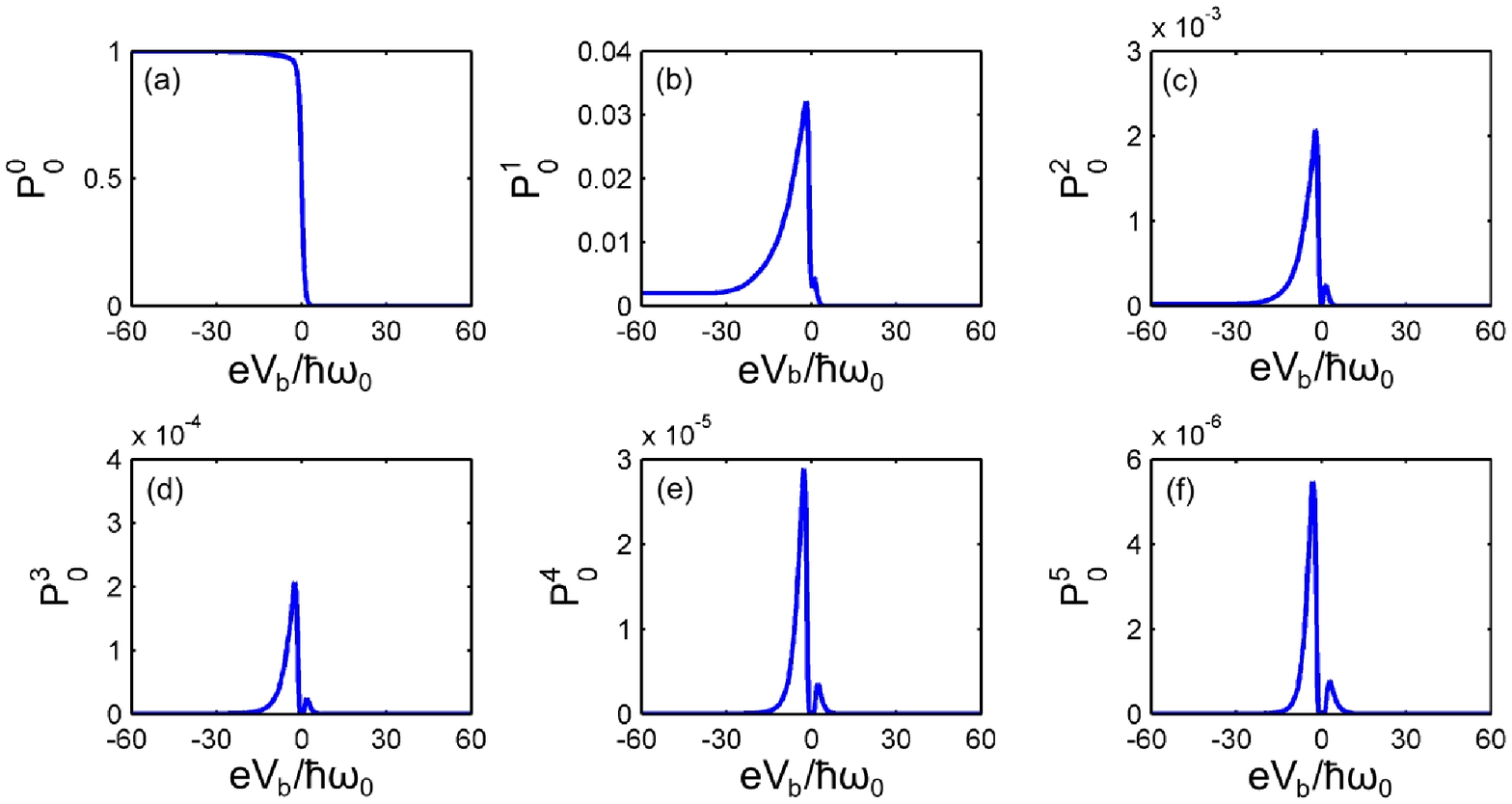,width=\columnwidth}}
\caption{(Color online). Populations $P^m_0$ as a function of normalized bias voltage for the 0-particle (a) vibronic ground state, (b) first excited state, (c) second excited state, (d) third excited state, (e)  fourth excited state, (f) fifth excited state, (g) sixth excited state, (h) seventh excited state, and (i) eighth excited state. The frequency of the driving field is $\omega \ll 1/\tau_{\rm max}$. The other parameters are the same as used in Figure~\ref{fig_s_t}.}
\label{fig_EVP_S}
\end{figure}

\begin{figure}[h]
\centerline{\psfig{figure=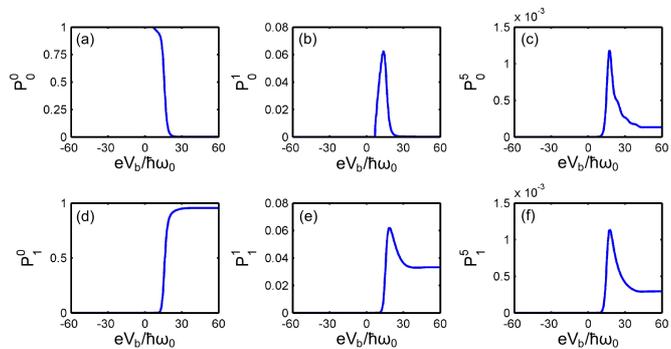,width=\columnwidth}}
\caption{(Color online). Population as a function of normalized bias voltage for the 0-particle (a) vibronic ground state, (b) first excited state, (c) fifth excited state and (d) 1-particle ground state, (e)  first excited state, (f) fifth excited state. The value of the gate voltage is $eV_\text{g}/\hbar\omega_0=8$, and the frequency of the driving field is $\omega \ll 1/\tau_{\rm max}$. The other parameters are the same as used in Figure~\ref{fig_s_t}.}
\label{fig_EVP_S8}
\end{figure}

In this section, we consider the limit ($\omega \to 0$) of DC-bias as a special case of the master equation presented in the previous section and compare the results. Even if the system still exhibits the bistable properties discussed in Section~\ref{sec:Lifetimes} (they are in fact not related to the sweeping time of the bias) the hysteretic behavior cannot be observed anymore. In Figure~\ref{fig_P_S}, we present the population of the electronic states for gate voltage $V_\text{g}=0$.
\begin{figure}[h]
\centerline{\psfig{figure=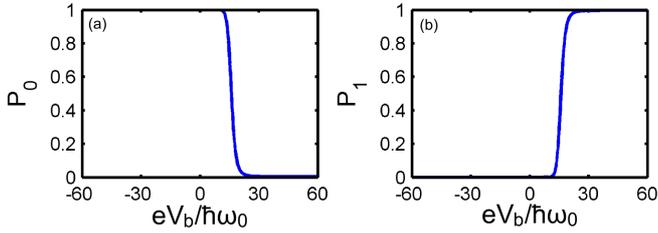,width=\columnwidth}}
\caption{(Color online). (a) Population of the 0-particle electronic state, (b) population of the 1-particle electronic state. The value of the gate voltage is $eV_\text{g}/\hbar\omega_0=8$, and the frequency of the driving field is $\omega \ll 1/\tau_{\rm max}$. The other parameters are the same as used in Figure~\ref{fig_s_t}.}
\label{fig_P_S8}
\end{figure}
At large negative bias the system is empty, while at large positive bias it is charged. The system makes transitions from the 0- to 1-particle state near zero bias.
Analogously, in Figure~\ref{fig_P_S8}, the population of electronic states as a function of normalized bias is depicted for gate voltage $eV_\text{g}/\hbar\omega_0=8$. Due to a finite $\varepsilon$, the transition $0\to 1$ occurs at positive bias voltages.
\begin{figure}[h]
\centerline{\psfig{figure=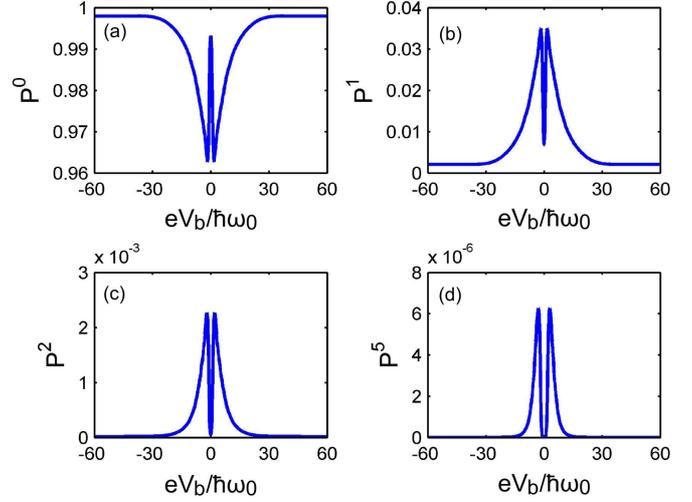,width=\columnwidth}}
\caption{(Color online). Populations as a function of normalized bias voltage for (a) vibronic ground state, (b) first excited state, (c) second excited state, and (d) fifth excited state. The frequency of the driving field is $\omega \ll 1/\tau_{\rm max}$. The other parameters are the same as used in Figure~\ref{fig_s_t}.}
\label{fig_VP_S}
\end{figure}
\begin{figure}[h]
\centerline{\psfig{figure=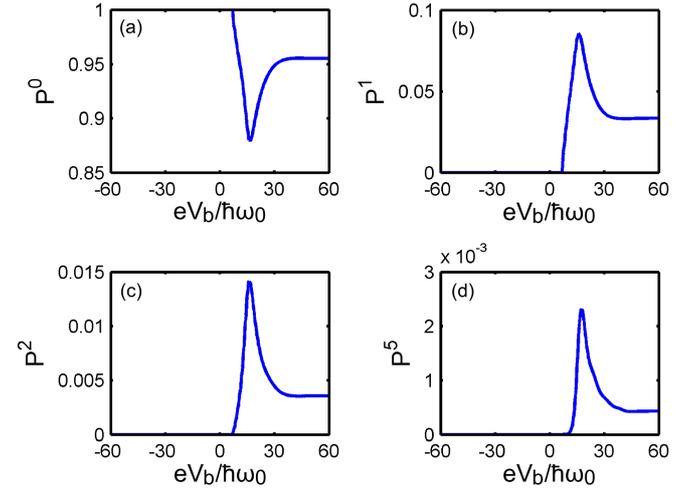,width=\columnwidth}}
\caption{(Color online). Population as a function of normalized bias voltage for (a) vibronic ground state, (b) first excited state, (c) second excited state, and (d) fifth excited state. The value of the gate voltage is $eV_\text{g}/\hbar\omega_0=8$, and the frequency of the driving field is $\omega \ll 1/\tau_{\rm max}$. The other parameters are the same as used in Figure~\ref{fig_s_t}.}
\label{fig_VP_S8}
\end{figure}
 Moreover, the populations of the vibronic states are sketched in Figure~\ref{fig_VP_S} for gate voltage $V_\text{g}=0$, which clearly shows that, for the considered parameters, only the vibronic ground state and first excited state are populated, whereas the populations of higher excited states are very small. This is in contrast to the non-stationary case where the excited states are highly populated (see Figure~\ref{fig_v_t}). In a similar way, the populations of the vibronic states for gate voltage $eV_\text{g}/\hbar\omega_0=8$ are presented in figure~\ref{fig_VP_S8} where higher excited states also get populated. Finally, in Figures~\ref{fig_EVP_S} and~\ref{fig_EVP_S8} we show the populations of the 0- and 1-particle vibronic states for gate voltages $V_\text{g}=0$ and $eV_\text{g}/\hbar\omega_0=8$, respectively, which basically provide the same information as mentioned before.

\subsection{$I-V$ characteristics for the DC-case}\label{subsec:StationaryCurrent}
\begin{figure}[h]
\centerline{\psfig{figure=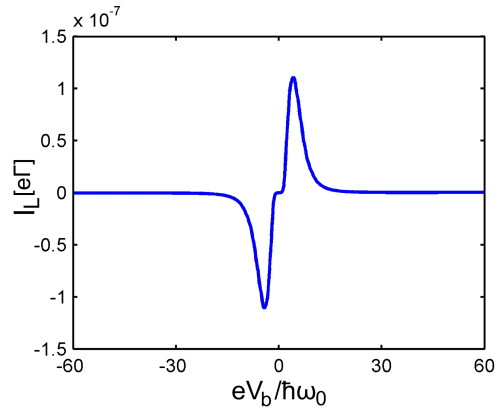,width=6.5cm}}
\caption{(Color online). Left current as a function of normalized bias. The frequency of the driving field is $\omega \ll 1/\tau_{\rm max}$. The other parameters are the same as used in Figure~\ref{fig_s_t}.}
\label{fig_I_s}
\end{figure}

\begin{figure}[h]
\centerline{\psfig{figure=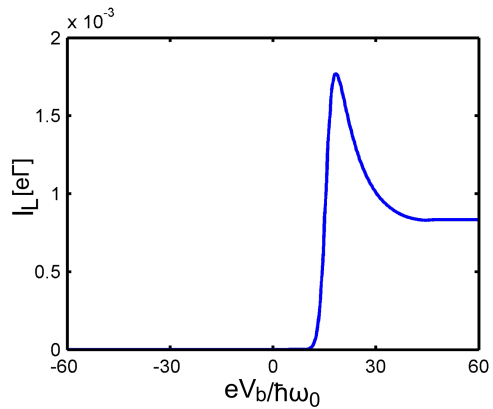,width=6.5cm}}
\caption{(Color online). Current as a function of normalized bias. The value of gate voltage is, $eV_\text{g}/\hbar\omega_0=8$. The other parameters are the same as used in Figure~\ref{fig_s_t}.}
\label{fig_I_S8}
\end{figure}

In the DC-case the analytical expression for the current remains the same as given by Eq.~\eqref{eq:C} taking into account a time independent bias.
Let us first discuss the situation when the 0- and 1-particle states are in resonance, $\varepsilon=\varepsilon_0-\varepsilon_\text{p}=0$ and $V_\text{g}=0$. In this particular case, an interesting behavior of the $I-V$ characteristics with two opposite current peaks around zero bias can be observed (see Figure~\ref{fig_I_s}).
In order to understand the mechanism of this process, we consider the source current which can be expressed in the form
\begin{align}\label{eq:CP}
I_\text{s}&={\rm{\Gamma}}_\text{s}\sum_{mm'}F_{mm'}\bigl\{f^+\bigl[\varepsilon+\hbar\omega_0\left(m'-m\right)-eV_b\bigr]P^m_0\nonumber\\&
-f^-\bigl[\varepsilon+\hbar\omega_0\left(m'-m\right)-eV_b\bigr]P^{m'}_1\bigr\}.
\end{align}
At $V_b=0$ only ground to ground state transitions are open and $P^0_0=P^0_1=\frac{1}{2}$. Hence, from Eq.\eqref{eq:CP} one deduces that in this region the current is zero. At large positive bias, i.e., $V_b\to \infty$, the current is zero because the system is in a 1-particle stable state and no new transition channel is available. For finite bias, the behavior of the Franck-Condon factor $F_{mm'}$ is of importance. In particular, it suffices to investigate the classically allowed transitions as determined by the Franck-Condon parabola~\cite{Nowack0506552,Herzberg45}. The minimum of the parabola is for $m=m'\sim\left(\frac{\lambda}{2}\right)^2$, i.e., $m,m'<\left(\frac{\lambda}{2}\right)^2$ transitions are exponentially suppressed. Moreover, $F_{mm'}$ attains the maximal values for $F_{mm'}=F_{m0}$ or $F_{mm'}=F_{0m'}$ and $m$ or $m'$ of the order of $\lambda^2$. Hence Fig.~\ref{fig_EVP_S} describes a threshold effect. The populations $P^m_1$ of the 1-particle states are mirror symmetric with respect to the bias inversion (not shown).
Analogously, we can analyze in the same way as above the current peak in Figure~\ref{fig_I_S8} which occurs at $eV_\text{b}\sim\varepsilon$ for gate voltage $eV_\text{g}/\hbar\omega_0 =8$.
\section{Conclusions}\label{sec:Summary}
In conclusion, we analyzed the quantum switching, bistability and memory effects in a single level system within the framework of the polaron model, where the electronic state is weakly coupled to metallic leads under AC-bias and strongly coupled to a vibrational mode. We showed that the bistability arises if the quantum switching between neutral and charged states involved is suppressed, e.g., due to Franck-Condon blockade.
In the case of an asymmetric junction, the neutral and charged states can be unstable at one polarity but stable at the other polarity of bias voltage. Under an appropriate choice of parameters, the stability regions of the two states overlap, which results in a bistable region in a certain interval of bias voltage. Taking into account non-stationary effects, in particular the interplay between time scales of variation of the external perturbation and the switching time of the system, we demonstrated electrically controlled hysteretic behavior of the system. Furthermore, we showed that vibronic states and average vibron energies also show hysteretic behavior like the ones shown by the population-voltage and current-voltage curves. At the end, we also discussed the case of a DC-bias.
In this case the population-voltage and current-voltage curves get single valued. Interestingly, one can observe current peaks in the $I-V$ characteristics of the system when given vibronic channels contribute to transport.
Moreover, we found that in the AC-case the vibronic excited states can be highly populated, while in the stationary case the population of the excited states is strongly decreased.
\section*{Acknowledgments}\label{sec:Acknowledgements}
Support of the DFG under the program GRK 1570 is acknowledged.
Abdullah Yar also acknowledges the support of Kohat University of Science \&
Technology, Kohat-26000, Khyber Pakhtunkhwa, Pakistan. We thank D. A. Ryndyk for useful discussions.
\appendix
\section{Calculation of the correlator $F_\alpha\left(t-t',\mu_0\right)$}\label{app:Correlator}
Here we calculate the correlation function $F_\alpha\left(t-t',\mu_0\right)$ of lead $\alpha$ in the wide band limit. From Eq.~\eqref{eq:ROH} we can write:
\begin{align}\label{eq:correlator}
F_\alpha\bigl(t-t',\mu_0\bigr)&=\sum_\kappa f_\alpha\bigl(\varepsilon_\kappa-\mu_0\bigr)e^{i\frac{\varepsilon_\kappa}{\hbar}(t-t')}\nonumber\\&=\int^{\infty}_{-\infty} {\rm{d}}\varepsilon D_\alpha f\bigl(\varepsilon-\mu_0\bigr)e^{i\frac{\varepsilon}{\hbar}(t-t')}\nonumber\\&=e^{i\frac{\mu_0}{\hbar}(t-t')}\int^{\infty}_{-\infty} {\rm{d}}\varepsilon D_\alpha f\bigl(\varepsilon\bigr)e^{i\frac{\varepsilon}{\hbar}(t-t')},
\end{align}
where $D_\alpha$ is the constant density of states of lead $\alpha$.
To simplify the above equation, we use the following relation: $f\left(\varepsilon\right)=\frac{1}{2}\left(1+\frac{e^{-\frac{\beta\varepsilon}{2}}-e^{\frac{\beta\varepsilon}{2}}}{e^{-\frac{\beta\varepsilon}{2}}
+e^{\frac{\beta\varepsilon}{2}}}\right)$. Hence
\begin{align}\label{eq:correlator2}
F_\alpha\bigl(t-t',\mu_0\bigr)=&\frac{D_\alpha}{2}e^{i\frac{\mu_0}{\hbar}(t-t')}\Biggl[\int^{\infty}_{-\infty} {\rm{d}}\varepsilon e^{i\frac{\varepsilon}{\hbar}(t-t')}\nonumber\\&+\int^{\infty}_{-\infty} {\rm{d}}\varepsilon \tanh\left(-\frac{\beta\varepsilon}{2}\right) e^{i\frac{\varepsilon}{\hbar}(t-t')}\Biggr].
\end{align}
The first term leads to the result
\begin{align}\label{eq:correlator4}
\int^{\infty}_{-\infty} {\rm{d}}\varepsilon e^{i\frac{\varepsilon}{\hbar}(t-t')}=2\pi\hbar\delta\bigl(t-t'\bigr).
\end{align}
The simplified form of the second part in ~\eqref{eq:correlator2} reads:
\begin{align}\label{eq:correlator5}
&\int^{\infty}_{-\infty} {\rm{d}}\varepsilon \tanh\left(-\frac{\beta\varepsilon}{2}\right) e^{i\frac{\varepsilon}{\hbar}(t-t')}\nonumber\\&=\int^{\infty}_{-\infty} {\rm{d}}\varepsilon\tanh\left(-\frac{\beta\varepsilon}{2}\right)\nonumber\\&\times\left\{\cos\left[\frac{\varepsilon}{\hbar}(t-t')\right]+i\sin\left[\frac{\varepsilon}
{\hbar}(t-t')\right]\right\}.
\end{align}
Due to symmetry the cosine component of the integral vanishes. One can further use the following relation:
\begin{multline}
\int^{\infty}_{0}{\rm{d}}x \sin(ax)\tanh\left(\frac{bx}{2}\right)=\frac{\pi}{b\tanh\bigl(\frac{\pi a}{b}\bigr)},\ \text{for}\ a,b\in\mathbb{R}.\label{eq:Formula}
\end{multline}
Using~\eqref{eq:Formula}, one can evaluate~\eqref{eq:correlator5} to be
\begin{align}\label{eq:correlator6}
\int^{\infty}_{-\infty} {\rm{d}}\varepsilon\tanh\left(-\frac{\beta\varepsilon}{2}\right)\sin\left[\frac{\varepsilon}{\hbar}(t-t')\right]
=-\frac{2\pi}{\beta\sinh\left[\pi\frac{(t-t')}{\hbar\beta}\right]}.
\end{align}
Putting all together, the correlation function gets the final form (in the wide band limit) as
\begin{multline}
F_\alpha\bigl(t-t',\mu_0\bigr)\\
=\pi\hbar D_\alpha e^{i\frac{\mu_0}{\hbar}(t-t')}\left\{\delta\bigl(t-t'\bigr)-\frac{i}{\hbar\beta\sinh\bigl[\pi\frac{(t-t')}{\hbar\beta}\bigr]}\right\}.\label{eq:correlator7}
\end{multline}
This function characterizes the correlation which exists on average between events where a lead electron is destroyed at time $t'$ and another is created at time $t$. It thus provides very important information about the time scales which control the relaxation dynamics of the leads.
\section{Evaluation of an integral}\label{app:Integral}
In order to solve Eq.~\eqref{eq:rhomarkov} and obtain the populations of the many-body states, one needs to evaluate the following integral:
\begin{align}\label{Integral}
I_\alpha=&\int^\infty_0 {\rm{d}}t''\left\{F\bigl[t'',\mu_\alpha(t)\bigr]e^{-\frac{i}{\hbar}{\rm{\Delta}} Et''}\right.\nonumber\\&\left.+F^\ast\bigl[t'',\mu_\alpha(t)\bigr]e^{\frac{i}{\hbar}{\rm{\Delta}} Et''}\right\}\nonumber\\&=2\int^\infty_0 {\rm{d}}t''Re\left\{F\left[t'',\mu_\alpha(t)\right]e^{-\frac{i}{\hbar}{\rm{\Delta}} Et''}\right\}.
\end{align}
Substituting the correlation function $F\bigl[t'',\mu_\alpha(t)\bigr]$ using Eq.~\eqref{eq:correlator7} in the above equation, we obtain
\begin{align}\label{eq:Integral}
I_\alpha=&\int^\infty_{-\infty} {\rm{d}}t''\pi\hbar D_\alpha\Biggl\{\cos\left[\frac{t''}{\hbar}\bigl(\mu_\alpha(t)-{\rm{\Delta}} E\bigr)\right]\delta(t'')\nonumber\\&+\frac{\sin\bigl[\frac{t''}{\hbar}\bigl(\mu_\alpha(t)-{\rm{\Delta}} E\bigr)\bigr]}{\hbar\beta\sinh\bigl(\pi\frac{t''}{\hbar\beta}\bigr)}\Biggr\}.
\end{align}
To simplify the above relation, one can use the following formula:
\begin{align}\label{eq:Formula2}
\int^\infty_{-\infty}{\rm{d}}x\frac{\sin(ax)}{\sinh(bx)}=\frac{\pi\tanh\left(\frac{a\pi}{2b}\right)}{b},\quad a,b\in\mathbb{R},\rm{and}\ b>0.
\end{align}
Using Eq.~\eqref{eq:Formula2}, we can write Eq.~\eqref{eq:Integral} as
\begin{align}\label{Integral3}
I_\alpha=\pi\hbar D_\alpha\left\{1+\tanh\left[\beta\frac{\mu_\alpha(t)-{\rm{\Delta}} E}{2}\right]\right\}.
\end{align}
After some calculations, we obtain
\begin{align}\label{Integral4}
I_\alpha=2\pi\hbar D_\alpha f\bigl[{\rm{\Delta}} E-\mu_\alpha(t)\bigr],
\end{align}
where $\mu_\alpha(t)=\mu_0+\rm{\rm{{\rm{\Delta}}}}\mu_\alpha(t)$.
\section{Evaluation of transition matrix elements of the electron operator}\label{app:Transitionrates}
To determine the transition rates, we need to calculate the matrix elements
\begin{align}\label{eq:transitionmatrixappendix1}
&\langle r\arrowvert \hat{d}\arrowvert s\rangle  =e^{-\frac{1}{2}\left|\lambda\right|^2}F\left(\lambda,m,m'\right),
\end{align}
where $\arrowvert r\rangle $ and $\arrowvert s\rangle $ represent the eigenstates given by Eq.~\eqref{eq:eigenstates}. The function $F(\lambda,m,m')$ determines the coupling between states with a different vibronic number of excitations with effective coupling $\lambda$ and is expressed as~\cite{mayrhoferEPJB2007,74PRB121403}
\begin{multline}
F(\lambda,m,m')=\\
\bigl[{\rm{\Theta}}(m'-m)\lambda^{m'-m}+{\rm{\Theta}}(m-m')\bigl(-\lambda^{*}\bigr)^{m-m'}\bigr]\\
\times\sqrt{\frac{m_{\text{min}}!}{m_{\text{max}}!}}\sum_{i=0}^{m_{\text{min}}}\frac{\bigl(-\left|\lambda\right|^{2}\bigr)^{i}}{i!(i+m_{\text{max}}-m_{\text{min}})!}
\frac{m_{\text{max}}!}{(m_{\text{min}}-i)!},\label{eq:Function}
\end{multline}
where $m_{\text{min}/\text{max}}=\text{min}/\text{max}(m,m')$.
The coefficient $F_{mm'}$ in Eqs.~\eqref{eq:tau0m} and~\eqref{eq:tau1mp} is defined as $F_{mm'}=e^{-\lambda^2}F^2(\lambda,m,m')$.

\section{Expression for the FC factor $F_{2m'}$}\label{app:F2m}
Using Appendix~\ref{app:Transitionrates} the expression for $F_{2m'}$ follows to be
\begin{align}\label{eq:F2m}
F_{2m'}=& e^{-\lambda^2}\biggl\{\frac{1}{2}\lambda^6-\frac{3}{2}\lambda^4+2\lambda^2\nonumber\\&+\frac{\lambda^{2(m'-2)}}{2!m'!}\left[m'^2-m'(1+2\lambda^2)+\lambda^4\right]^2\biggr\}.
\end{align}

\end{document}